\documentclass[twocolumn,showpacs]{revtex4-1}
\usepackage{newlfont}
\usepackage{times}
\usepackage[below]{placeins}
\usepackage{flafter}
\usepackage{amsfonts}
\usepackage{amsmath}
\usepackage{amssymb}
\usepackage[american]{babel}
\usepackage{graphicx}
\usepackage{subfigure}
\usepackage{multirow}
\usepackage{url}
\usepackage{color}
\usepackage{graphicx}
\usepackage{mathtools}
\usepackage{soul,xcolor}
\setstcolor{blue}
\allowdisplaybreaks
\newcommand{\ket}[1]{\big| #1 \big\rangle}
\newcommand{\bra}[1]{\big\langle #1 \big|}
\newcommand{\braket}[2]{\big\langle #1 \big| #2 \big\rangle}                 % < | >
\begin{document}

\title{Staggered Quantum Walks with Hamiltonians}
\author{R. Portugal}
\affiliation{Laborat\'orio Nacional de Computa\c c\~ao Cient\'\i fica (LNCC), Petr\'opolis, RJ, Brazil}
\author{M. C. de Oliveira }\
\affiliation{Instituto de F\'\i sica Gleb Wataghin, Universidade Estadual de Campinas, Campinas, SP, Brazil}
\author{J. K. Moqadam}
\affiliation{Instituto de F\'\i sica Gleb Wataghin, Universidade Estadual de Campinas, Campinas, SP, Brazil}

%\author{Collaboration LNCC-Unicamp}%%R. Portugal\\ %%%\mbox{$^{1}$} 
%\affiliation{LNCC-Unicamp}
%{\small National Laboratory of Scientific Computing - LNCC} \\
%%%{\small $^1$Laborat\'{o}rio Nacional de Computa\c{c}\~{a}o Cient\'{i}fica - LNCC} \\
%{\small Av. Get\'{u}lio Vargas 333, 25651-075, Petr\'{o}polis, RJ, Brazil}\\

\begin{abstract}
Quantum walks are recognizably useful for the development of new quantum algorithms, as well as for the investigation of several physical phenomena in quantum systems. Actual implementations of quantum walks face technological difficulties similar to the ones for quantum computers, though. Therefore, there is a strong motivation to develop new quantum-walk models which might be easier to implement. In this work, we present an extension of the staggered quantum walk model that is fitted for physical implementations in terms of time-independent Hamiltonians. We demonstrate that this class of quantum walk includes the entire class of staggered quantum walk model, Szegedy's model, and an important subset of the coined model.
\end{abstract}

\date{\today}
\maketitle

\section{Introduction}

Coined quantum walks (QWs) on graphs were firstly defined in Ref.~\cite{Aharonov:2000} and have been extensively analyzed in the literature~\cite{Ven12,Kon08,Kendon:2007,Portugal:Book,Manouchehri2014}. Many experimental proposals for the QWs were given previously~\cite{Travaglione2002,sanders2003quantum,MPO15}, with some actual experimental implementations  performed in Refs.~\cite{Karski2009,Zahringer2010,Schreiber2010}. The key feature of the coined QW model is to use an internal state that determines possible directions that the particle can take under the action of the shift operator (actual displacement through the graph). Another important feature is the alternated action of two unitary operators, namely, the coin and shift operators. Although all discrete-time QW models have the ``alternation between unitaries'' feature, the coin is not always necessary because the evolution operator can be defined in terms of the graph vertices only, without using an internal space as, for instance, in Szegedy's model~\cite{Szegedy:2004} or in the ones described in Refs.~\cite{AFC08,BDEPT16}.

More recently, the staggered quantum walk (SQW) model was defined in Refs.~\cite{PSFG15,Por16b}, where a recipe to generate unitary and Hermitian local operators based on the graph structure was given. The evolution operator in the SQW model is a product of local  operators \footnote{An operator is called local in the context of quantum walks if its action on a particle that is on a vertex $v$ moves the particle to the neighborhood of $v$, but not further away.}.  The SQW model contains a subset of the coined QW class of models~\cite{Aharonov:2000}, as shown in Ref.~\cite{Por16}, and the entire Szegedy model~\cite{Szegedy:2004} class.

Although covering a more general class of quantum walks, there is a restriction on the local evolution operations in the SQW demanding Hermiticity besides unitarity. This severely compromises the possibilities for actual implementations of SQWs on physical systems because the unitary evolution operators, given in terms of time-independent Hamiltonians having the form $U_0=\textrm{e}^{i\theta_0 H_0}$, are non-Hermitian in general. To have a model, that besides being powerful as the SQW, to be also fitted for practical physical implementations, it would be necessary to relax on the Hermiticity requirement for the local unitary operators.
\begin{figure}[ht!] 
\centering
\includegraphics[trim = 27mm 60mm 0mm 0mm, clip=true, scale=0.25]{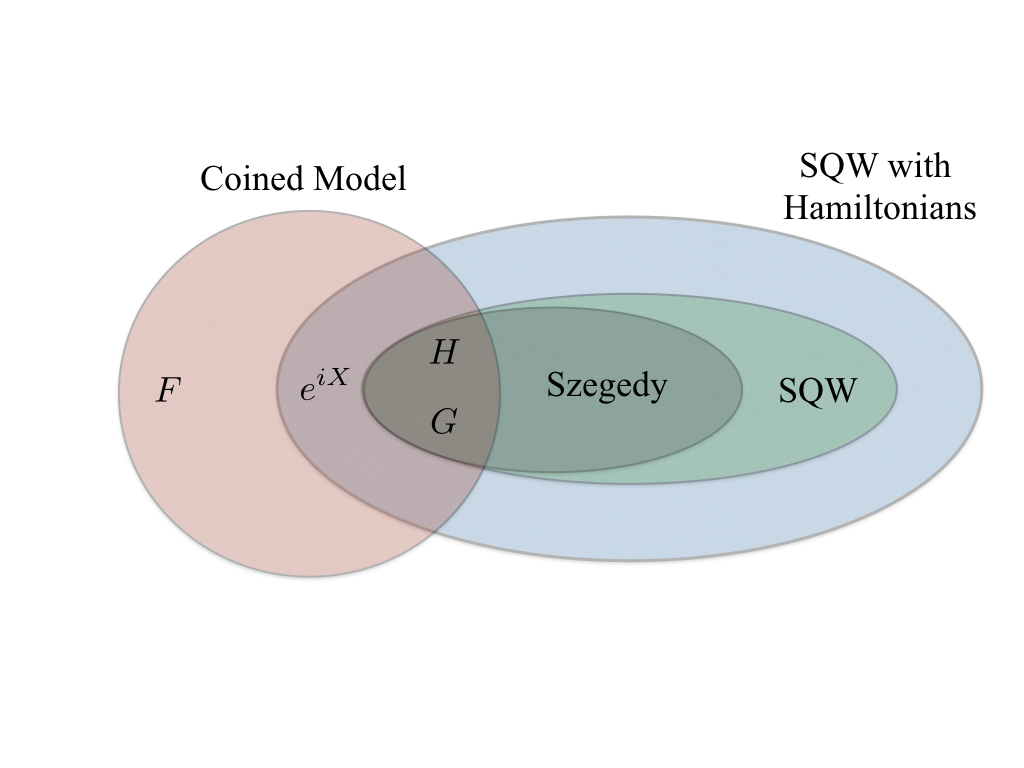}
\caption{Classes of discrete-time QWs. The set of SQWs with Hamiltonians proposed in this work includes the entire set of standard SQWs, which includes the entire Szegedy's model. The coined model is partially represented in the SQW model. $H$, $G$, and $F$ stand for flip-flop coined QWs with Hadamard, Grover, and Fourier coins.} 
\label{SQWWH_graph1}
\end{figure}

In this work, we propose an extension of the SQW model employing non-Hermitian local operators. The concatenated evolution operator has the form
$$U\,=\,\textrm{e}^{i\theta_1 H_1}\textrm{e}^{i\theta_0 H_0},$$
where $H_0$ and $H_1$ are unitary and Hermitian, $\theta_0$ and $\theta_1$ are general angles representing specific systems' energies and time intervals (divided by the Planck constant $\hbar$). The standard SQW model is recovered when $\theta_0=\pm \pi/2$ and $\theta_1=\mp \pi/2$. With this modification, SQW with Hamiltonians encompasses the standard SQW model and includes new coined models. Besides, with the new model, it is easier to devise new experimental proposals such as the one described in Ref.~\cite{coinless}.

Fig.~\ref{SQWWH_graph1} depicts the relation among the discrete-time QW models. Szegedy's model is included in the standard SQW model class, which itself is a subclass of the SQW model with Hamiltonians. Flip-flop coined QWs that are in Szegedy's model are also in the SQW model. Flip-flop coined QWs using Hadamard $H$ and Grover $G$ coins, as represented in Fig.~\ref{SQWWH_graph1}, are examples. There are coined QWs, which are in  the SQW model with Hamiltonians general class, but not in the standard SQW model, as for example, the one-dimensional QWs with coin $\textrm{e}^{i\theta X}$, where $X$ is the Pauli matrix $\sigma_X$, with  angle $\theta$ not a multiple of $\pi/2$. Those do not encompass all the possible coined QW models, as there are flip-flop coined QWs, which although being built with non-Hermitian unitary evolution, cannot be put in the SQW model with Hamiltonians. For instance, when the Fourier coin $F$ is employed, where $F_{ij}=\omega^{ij}$ and $\omega=\exp(2i\pi/N)$, being $N$ the Hilbert space dimension.

The structure of this paper is as follows. In Sec.~\ref{sec2}, we describe how to obtain the evolution operator of the SQW with Hamiltonians on a generic simple undirected graph. In Sec.~\ref{sec3}, we calculate the wave function using the Fourier analysis for the one-dimensional lattice and the standard deviation of the probability distribution.   In Sec.~\ref{sec4}, we characterize which coined QWs are included in the class of SQWs with Hamiltonians. Finally, in Sec.~\ref{sec5} we draw our conclusions.

\section{The evolution operator}\label{sec2}
Let $\Gamma(V,E)$ be a simple undirected graph with vertex set $V$ and edge set $E$.  A tessellation of $\Gamma$ is a partition of $V$ so that each element of the partition is a clique. A clique is a subgraph of $\Gamma$ that is complete. An element of the partition is called a polygon. The tessellation covers all vertices but not necessarily all edges. Let ${\cal H}$ be the Hilbert space spanned by the computational basis $\big\{\ket{v}:v\in V\big\}$, that is, each vertex $v$ is associated with a vector $\ket{v}$ of the canonical basis. Each polygon spans a subspace of the ${\cal H}$, whose basis comprises the vectors of the computational basis associated with the vertices in the polygon. Let $m$ be the number of polygons and let $\alpha_k$ be a polygon for some $0\le k< m$. A unit vector \textit{induces} polygon $\alpha_k$ if the following two conditions are fulfilled: First, the vertices of $\alpha_k$ is a clique in $\Gamma$. Second, the vector has the form
\begin{equation} \label{alpha_k}
 \ket{\alpha_k} \,=\,  \sum_{v\in V} a_{k,v} \ket{v},
\end{equation}
so that $a_{k,v}\neq 0$ for $v\in \alpha_k$ and $a_{k,v}=0$ otherwise. The simplest choice is the uniform superposition given by $a_{k,v}=1/\sqrt{\left|\alpha_k\right|}$ for $v\in \alpha_k$.

There is a recipe to build a unitary and Hermitian operator associated with the tessellation, when we use the following structure: 
\begin{equation}
  H_0 \,=\, 2\sum_{k=0}^{m-1} \ket{\alpha_k}\bra{\alpha_k} - I. \label{H_0}
\end{equation}
$H_0$ is unitary because the polygons are non-overlapping, that is, $\braket{\alpha_k}{\alpha_{k'}}=\delta_{k k'}$ for $0\le k,k' < m$.  $H_0$ is Hermitian because it is a sum of Hermitian operators. Then, $H_0^2=I$. An operator of this kind is called an \textit{orthogonal reflection} of graph $\Gamma$. Each $\alpha_k$ induces a polygon and we say that $H_0$ induces the tessellation.

The idea of the staggered model is to define a second operator that must be independent of $H_0$. Define a second tessellation by making another partition of $\Gamma$ with polygons $\beta_k$ for $0\le k<n$, where $n$ is the number of polygons. For each polygon $\beta_k$, define unit vectors
\begin{equation} \label{beta_k}
 \ket{\beta_k} \,=\,  \sum_{v\in V} b_{k,v} \ket{v},
\end{equation}
so that $b_{k,v}\neq 0$ for $v\in \beta_k$ and $b_{k,v}=0$ otherwise.  The simplest choice is the uniform superposition given by $b_{k,v}=1/\sqrt{\left|\beta_k\right|}$ for $v\in \beta_k$. Likewise, define
\begin{equation}
  H_1 \,=\, 2\sum_{k=0}^{n-1} \ket{\beta_k}\bra{\beta_k} - I. \label{H_1}
\end{equation}
$H_1$ is an orthogonal reflection.

To obtain the evolution operator we demand that the union of tessellations $\alpha$ and $\beta$ should cover the edges of $\Gamma$, where tessellation $\alpha$ is the union of polygons $\alpha_k$ for $0\le k<m$ and tessellation $\beta$ is the union of polygons $\beta_k$  for $0\le k<n$. This demand is necessary because edges that do not belong to the tessellation union can be removed from the graph without changing the dynamics.

The standard SQW dynamics is given by the evolution operator $U = H_1 H_0$ where the unitary and Hermitian operators $H_0$ and $H_1$ are constructed as described in Eqs.~(\ref{H_0})~and~(\ref{H_1}). However, such graph-based construction of the operators does not correspond, in general, to the evolution of the real physical systems which are unitary but non-Hermitian instead. Actually, the unitary and non-Hermitian operators do not have a nice representation as in Eqs.~(\ref{H_0})~and~(\ref{H_1}). In the following, we introduce and analyze a method for constructing ``physical evolutions'' using the graph-based unitary and Hermitian operators.

We define the staggered QW model with Hamiltonians by the evolution operator
\begin{equation}\label{U}
	U\,=\,\textrm{e}^{i\theta_1 H_1}\textrm{e}^{i\theta_0 H_0},
\end{equation}
where $\theta_0$ and $\theta_1$ are angles. $U$ can be written as
\begin{equation}
U\,=\,\left(\cos\theta_1\,I+i\sin\theta_1\,H_1\right)\left(\cos\theta_0\,I+i\sin\theta_0\,H_0\right).
\end{equation}
%\begin{equation}
%U\,=\,\left(\cos\theta_0\cos\theta_1\,I- \sin\theta_0\sin\theta_1\,H_1\,H_0\right)+
%i\left(\sin\theta_0\cos\theta_1 H_0+\cos\theta_0\sin\theta_1 H_1\right).
%\end{equation}
The standard SQW model is obtained when $\theta_0=\pm\pi/2$ and $\theta_1=\mp\pi/2$.

The staggered QW model with Hamiltonians is characterized by two tessellations and the angles $\theta_0$ and $\theta_1$. The evolution operator is the product of two \textit{local} unitary operators. Local in the sense discussed before, that is, if a particle is on vertex $v$, it will move to the neighborhood of $v$ only. Some graphs are not 2-tessellable as discussed in Ref.~\cite{Por16b}. In this case, we have to use more than two tessellations until covering all edges and Eq.~(\ref{U}) must be extended accordingly.
\begin{figure}[ht] 
\centering
\includegraphics[trim = 47mm 95mm 0mm 0mm, clip=true, scale=0.3]{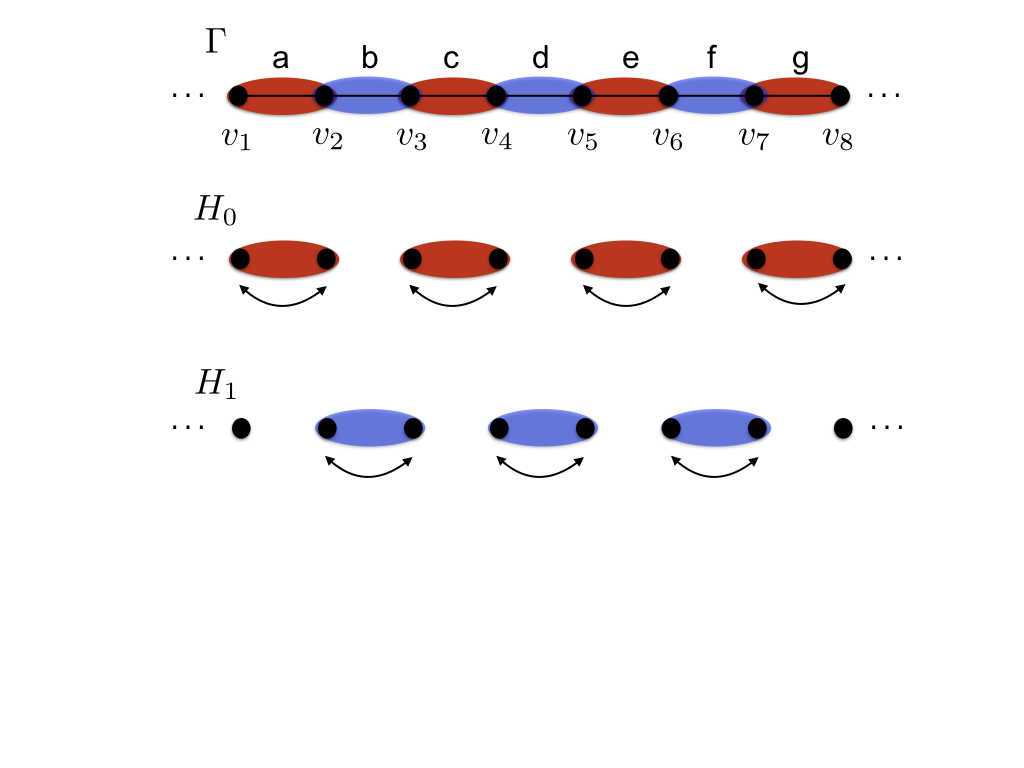}
\caption{One-dimensional lattice, with the two possible tesselations $\alpha$ (red), and $\beta$ (blue), and the local Hamiltonians $H_0$ and $H_1$ implementing it.} 
\label{1dlattice}
\end{figure}

\section{One-dimensional SQW with Hamiltonians}\label{sec3}
One of the simplest example of a SQW model with Hamiltonians is the one-dimensional lattice (or chain) as in Fig.~\ref{1dlattice}. If we wish to use the minimum number of tessellations that cover all vertices and edges, the only choice are the two tesselations represented in the figure and correspond to two alternate interactions between first neighbors. Therefore the evolution operator in the one-dimensional case with $\theta_0=\theta_1=\theta$ is given by
\begin{equation}
	U=\textrm{e}^{i\theta H_1}\textrm{e}^{i\theta H_0},\label{unit}
\end{equation}
where 
\begin{eqnarray}
	H_0 &=& 2\sum_{x=-\infty}^\infty \ket{u_x^{0}}\bra{u_x^{0}}-I,\label{h0} \\
	H_1 &=& 2\sum_{x=-\infty}^\infty \ket{u_x^{1}}\bra{u_x^{1}}-I, \label{h1}
\end{eqnarray}
and
\begin{eqnarray}
\left|u_{x}^{{0}}\right\rangle  & = & \cos\frac{\alpha}{2}\left|2x\right\rangle +\textrm{e}^{i\phi_0}\sin\frac{\alpha}{2}\left|2x+1\right\rangle,\label{eq:GenUAL_0}\\
\left|u_{x}^{{1}}\right\rangle  & = & \cos\frac{\beta}{2}\left|2x+1\right\rangle +\textrm{e}^{i\phi_1}\sin\frac{\beta}{2}\left|2x+2\right\rangle.\label{eq:GenUAL_1}
\end{eqnarray}
For the sake of simplicity, we choose $\alpha$ and $\beta$ to be independent from $x$. $U$ is defined on Hilbert space ${\cal H}$, whose computational basis is $\big\{\ket{x}:x\in\mathbb{Z}\big\}$. 

While the diagonal forms of the Hamiltonians~(\ref{h0}) and~(\ref{h1}) with $(+1)$-eigenvectors~(\ref{eq:GenUAL_0}) and~(\ref{eq:GenUAL_1}), respectively, are more appropriate to the QW related computations, one cannot immediately see the connections to interactions energies that they usually represent. For actual implementations, it is more convenient to write down it in terms of bosonic operators as
\begin{eqnarray}
	H_0 &=& \sum_{j} \frac{\omega_j}{2} a_j^\dagger a_j+ \lambda \sum_{j\;\textrm{odd}} \left(a_j^\dagger a_{j+1}+a_{j} a_{j+1}^\dagger\right),\label{h00} \\
    H_1 &=& \sum_{j} \frac{\omega_j}{2} a_j^\dagger a_j+ \lambda \sum_{j\;\textrm{even}} \left(a_j^\dagger a_{j+1}+a_{j} a_{j+1}^\dagger\right). \label{h11}
\end{eqnarray}
In that form the first term represents the occupations of each site and the second one represents hopping Hamiltonians. Note that since the QW models considered here are single particles quantum walks, the corresponding picture in terms of Hamiltonians (\ref{h00}) and (\ref{h11})  implementation  is to consider a single excitation in the encoding physical system. The joint Hamiltonian $H_0+H_1$ describes a large number of physical systems, from cold atoms trapped in optical lattices~\cite{Greiner2002, Jaksch200552} to a linear array of electromechanical resonators~\cite{1508.06984}. However the alternated action of the two local unitary operators in~(\ref{unit}) requires that the Hamiltonians $H_0$ and $H_1$ be applied independently. This requires a more involved process of alternating interactions in the system, which demands an  external control particular to each physical system. A proposal on how to implement it in a one dimensional array of coupled superconducting transmission line resonators is discussed elsewhere~\cite{coinless}.

\begin{figure}[ht!] 
\centering
\includegraphics[trim=50 516 250 50,clip,scale=0.7]{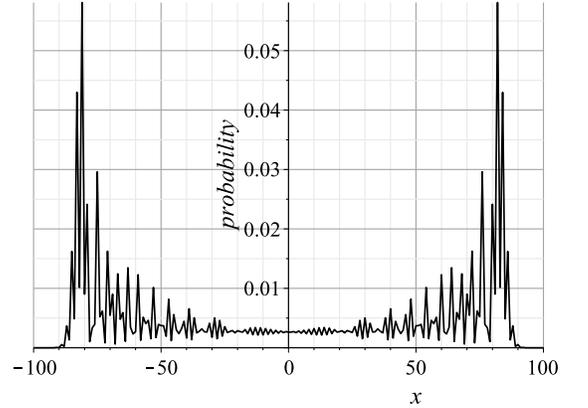}
%%trim=left down right up
\caption{Probability distribution after 60 steps with parameters $\theta=\pi/4$, $\alpha=\beta=\pi/2$, $\phi_0=\phi_1=0$, and initial condition $(\ket{0}+\ket{1})/\sqrt 2$.}
\label{fig:SQWWH_graph2}
\end{figure}

To start our analysis, in Fig.~\ref{fig:SQWWH_graph2} we show the probability distribution for the 1d SQW with Hamiltonians (\ref{h0}) and (\ref{h1}) after 60 steps with parameters $\theta=\pi/4$, $\alpha=\beta=\pi/2$, and $\phi_0=\phi_1=0$. The initial condition assumed was $(\ket{0}+\ket{1})/\sqrt 2$. A quantum walk with those parameters was analyzed by Ref.~\cite{Patel05}. Note the typical profile, which is similar to the coined QW, but certainly not to the continuous-time QW~\cite{Portugal:Book}.

\subsection{Fourier analysis}

%%\textcolor{red}{Why do we need this subsection? We should explain it.}

%
In order to find the spectral decomposition of the evolution operator, 
we perform a basis change that takes advantage of the system symmetries. Let us define the Fourier basis by the vectors
\begin{eqnarray}
\ket{\tilde{\psi}_{k}^{\textrm{0}}} & = & \sum_{x=-\infty}^{\infty}\textrm{e}^{-2xki}\ket{2x},\label{eq:Pie}\\
\ket{\tilde{\psi}_{k}^{\textrm{1}}} & = & \sum_{x=-\infty}^{\infty}\textrm{e}^{-(2x+1)ki}\ket{2x+1},\label{eq:Pio}
\end{eqnarray}
where $k\in[-\pi,\pi]$. For a fixed $k$, those vectors define a plane that
is invariant under the action of the evolution operator, which is confirmed by the following results:
\begin{eqnarray}
U\ket{\tilde{\psi}_{k}^{\textrm{0}}} & = &A\ket{\tilde{\psi}_{k}^{\textrm{0}}}+ B\ket{\tilde{\psi}_{k}^{\textrm{1}}},\label{eq:Hpsitilde1}\\
U\ket{\tilde{\psi}_{k}^{\textrm{1}}} & = & -B^*\ket{\tilde{\psi}_{k}^{\textrm{0}}}+A^*\ket{\tilde{\psi}_{k}^{\textrm{1}}},\label{eq:Hpsitilde2}
\end{eqnarray}
where 
\begin{eqnarray}
A & = &\sin^{2} \theta \left( \cos \alpha \cos
 \beta -\sin \alpha \sin \beta\,
 {{\textrm e}^{i \left( {\phi_0}+{\phi_1}+2k \right)}}\right)  \nonumber\\ 
&&  + \cos^{2}\theta + i\sin \theta \cos \theta
  \left( \cos \alpha -\cos \beta \right),\label{eq:A}\\
B & = & \sin\theta \sin\alpha  \left( i\cos\theta -\sin\theta \cos\beta \right) 
{{\textrm e}^{i({\phi_0}+k)}}\nonumber\\ 
&&+\sin\theta \sin\beta  \left( i\cos\theta -\sin\theta \cos\alpha \right)
 {{\textrm e}^{-i(\phi_1+k)}}
.\label{eq:B}
\end{eqnarray}
The analysis of the dynamics can be reduced to a two-dimensional subspace of ${\cal H}$ by defining a reduced evolution operator 
\begin{equation}
U_{\textrm{RED}}^{(k)}=\left[\begin{array}{cc}
 A & - B^* \\
 B &  A^*
\end{array}\right].
\end{equation}
$U_{\textrm{RED}}^{(k)}$ is unitary since $A\, A^{*}+B\, B^{*}=1.$ A vector in this subspace is mapped to Hilbert space ${\cal H}$ after multiplying its first entry by $\ket{\tilde{\psi}_{k}^{\textrm{0}}}$ and its second entry by $\ket{\tilde{\psi}_{k}^{\textrm{1}}}$. 

The eigenvalues of $U_{\textrm{RED}}^{(k)}$ (the same of $U$) are $\textrm{e}^{\pm i\lambda}$, where
\begin{equation}\label{eq:costheta}
\cos\lambda=\frac{A+A^{*}}{2}.
\end{equation}
Note that $A$ in (\ref{eq:A}) depends on $k$, as well as others parameters. The non-trivial eigenvectors of $U_{\textrm{RED}}^{(k)}$ are 
\begin{equation}
\frac{1}{\sqrt{C^{\pm}}}\left(\begin{array}{c}
-B^{*}\\
\textrm{e}^{\pm i\lambda}-A
\end{array}\right),\label{eq:eigenvec}
\end{equation}
where 
\begin{equation}
C^{\pm}=\sin\lambda\,\big(2\sin\lambda\pm i\,(A-A^{*})\big).
\end{equation}
The eigenvectors of the evolution operator $U$ associated with eigenvalues $\textrm{e}^{\pm i\lambda}$
are 
\begin{equation}
\ket{v_{k}^{\pm}}=\frac{1}{\sqrt{C^{\pm}}}\left(-B^{*}\ket{\tilde{\psi}_{k}^{\textrm{0}}}+(\textrm{e}^{\pm i\lambda}-A)\ket{\tilde{\psi}_{k}^{\textrm{1}}}\right),
\label{eq:Uev}
\end{equation}
and we can write
\begin{equation}
U=\int_{-\pi}^{\pi}\frac{\textrm{d}k}{2\pi}\left(\textrm{e}^{i\lambda}\ket{v_{k}^{+}}\bra{v_{k}^{+}}+\textrm{e}^{-i\lambda}\ket{v_{k}^{-}}\bra{v_{k}^{-}}\right).\label{eq:U}
\end{equation}

\begin{figure}[!h] 
\centering
\includegraphics[trim=50 490 250 50,clip,scale=0.7]{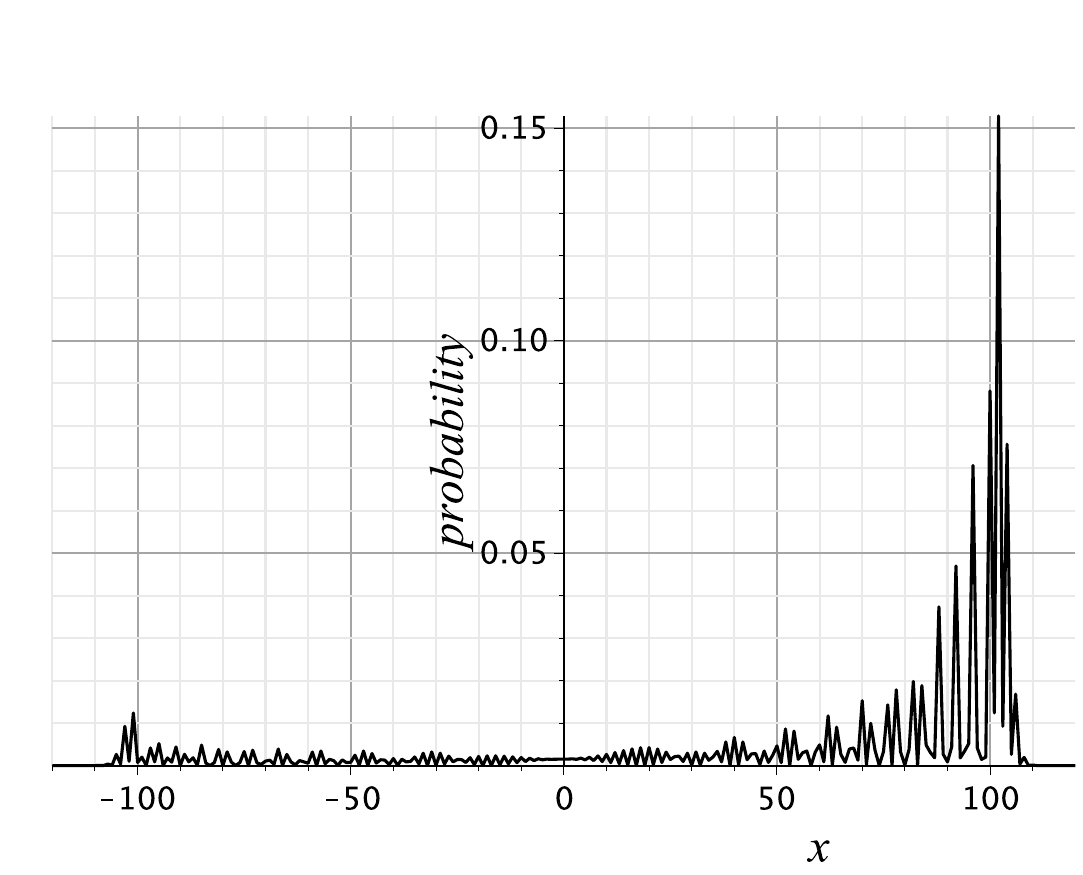}
%%trim=left down right up
\caption{Probability distribution after 60 steps with parameters $\theta=\pi/3$, $\alpha=\beta=\pi/2$, $\phi_0=\phi_1=0$, and initial condition $\ket{0}$.}
\label{fig:SQWWH_graph2a}
\end{figure}

If we take $\ket{\psi(0)}=\ket{0}$ as the initial condition, the quantum walk state at time $t$ is given by
\begin{eqnarray}
\ket{\psi(t)} & = & \sum_{x=-\infty}^{\infty}\left(\psi_{2x}(t)\,\ket{2x}+\psi_{2x+1}(t)\,\ket{2x+1}\right),
\end{eqnarray}
where 
\begin{equation}
\psi_{2x}(t)=\int_{-\pi}^{\pi}\frac{\textrm{d}k}{2\pi}\,|B|^{2}\left(\frac{\textrm{e}^{i(\lambda t-2kx)}}{C^{+}}+\frac{\textrm{e}^{-i(\lambda t+2kx)}}{C^{-}}\right),\label{eq:psike}
\end{equation}
and 
\begin{equation}
\psi_{2x+1}(t)=\int_{-\pi}^{\pi}\frac{\textrm{d}k}{2\pi}\,\frac{B\sin\lambda t}{\sin\lambda}\,\textrm{e}^{-(2x+1)ki}.\label{eq:psiko}
\end{equation}
The probability distribution is obtained after calculating $p_{2x}(t)=\left|\psi_{2x}(t)\right|^2$ and $p_{2x+1}(t)=\left|\psi_{2x+1}(t)\right|^2$.  The probability distribution would not be symmetric in this case (localized initial condition), as can be seen in Fig.~\ref{fig:SQWWH_graph2a}. Those results extend the corresponding ones obtained in Ref.~\cite{PBF15}.

\subsection{Standard deviation}\label{sec3a}
The results of Ref.~\cite{SPB15} can be extended in order to include parameter $\theta$ of the SQW with Hamiltonians. The asymptotic expression for the odd moments with initial condition $\ket{\psi(0)}=\ket{0}$ is
\begin{equation}\label{eq:oddm}
\left< x^{2n-1}\right>_t = \frac{t^{2n-1}}{4\pi}\int_{-\pi}^{\pi}\left[\frac{A-A^*}{i\sin\lambda}\right]^{2n}dk+O(t^{2n-2}),
\end{equation}
and for the even moments is
\begin{equation}\label{eq:evenm}
\left< x^{2n}\right>_t = 2t\left< x^{2n-1}\right>_t +O(t^{2n-1}).
\end{equation} 
The square of the standard deviation is
\begin{equation}
\sigma^2 = (2t-\left< x\right>_t)\left< x\right>_t.
\end{equation}
For $\alpha=\beta\le \pi/2$ and $\phi_0=\phi_1=0$, it simplifies asymptotically to
\begin{equation}
\sigma^2\,=\, 4\,\sqrt {1- \sin^{2} \theta\sin^{2}\alpha } 
\left(1- \sqrt {1- \sin^{2}\theta \sin^{2}\alpha } \right) {t^2}.
\end{equation}
\begin{figure}[ht!] 
\centering
\includegraphics[trim=30 430 0 158,clip,scale=0.5]{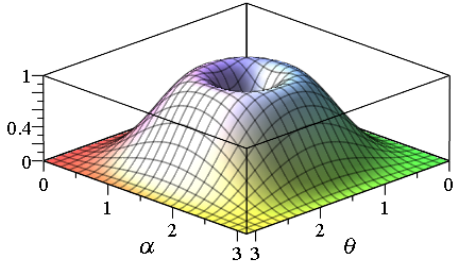}
%%trim=left down right up
\caption{Plot of $\sigma^2/t^2$ as function of $\theta$ and $\alpha$. The
value of $\sigma^2/t^2$ at the center of the plot is zero.}
\label{fig:SQWWH_graph5}
\end{figure}

Fig.~\ref{fig:SQWWH_graph5} shows the plot of $\sigma^2/t^2$ as function of $\theta$ and $\alpha$. The maximum value of $\sigma^2/t^2$ is $1$, which is achieved for the points on a circle with center at $(\pi/2,\pi/2)$ and radius $\pi/6$, for instance, $\sigma=t+O(1)$ when $\theta=\pi/3$ and $\alpha=\beta=\pi/2$. When $\alpha=\beta=\pi/2$ and $\phi_0=\phi_1=0$, $H_0$ is the direct sum of Pauli $X$ matrices
\begin{equation}
H_0= \left[ 
\begin{array}{ccc} 
\ddots&&0\\ 
&X&\\
0&&\ddots
\end{array} 
\right] 
,
\end{equation}
likewise $H_1$, with a diagonal shift of one entry.

\section{Coined QWs that are in the SQW with Hamiltonians}\label{sec4}

Any flip-flop coined QW on a graph $\Gamma(V,E)$ with a coin operator of the form $\textrm{e}^{i\theta_0 H_0}$, where $H_0$ is an orthogonal reflection of $\Gamma$,  is equivalent to a SQW with Hamiltonians on a larger graph $\Gamma'(V',E')$. The procedure to obtain $\Gamma'(V',E')$ is described in Ref.~\cite{Por16}. We briefly review it in the next paragraph.
% To show this we have to describe $\Gamma'(V',E')$ and the tessellations.

\begin{figure}[!h] 
\centering
\includegraphics[scale=0.26]{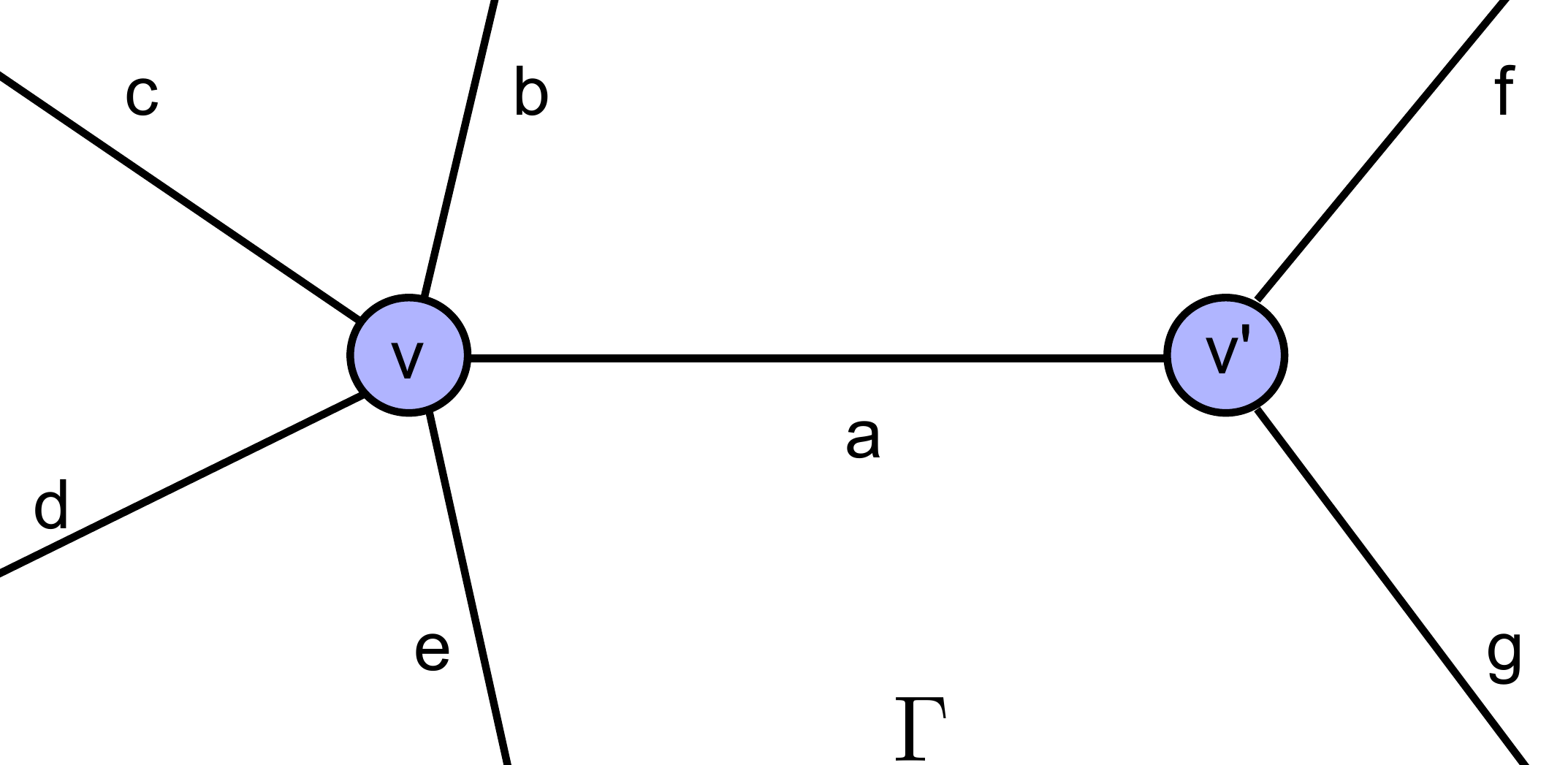}
\caption{Example of part of a graph showing a degree-5 vertex $v$ and a degree-3 vertex $v'$. Edge $(v,v')$ has label $a$. The other edges have labels $b$ to $g$. }
\label{fig:SQWWH_graph6a}
\end{figure}

\begin{figure}[!h] 
\centering
\includegraphics[scale=1.32]{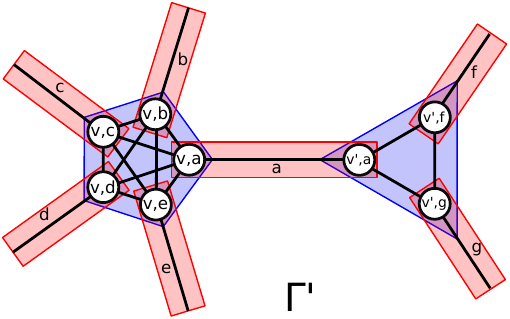}
\caption{To define a SQW that is equivalent to a flip-flop coined QW on the graph of Fig.~\ref{fig:SQWWH_graph6a}, one has to replace a degree-$d$ vertex by a $d$-clique. The vertex labels of the enlarged graph have the form ``$v,j$'', where $v$ is the label of the vertex in the original graph and $j$ is the edge incident on $v$.
}
\label{fig:SQWWH_graph6b}
\end{figure}

Let $0,...,|V|-1$ be the vertex labels and let $0,...,|E|-1$ be the edge labels of graph $\Gamma$. The action of the flip-flop shift operator on vectors of the computational basis associated with $\Gamma$ is
\begin{equation}
	S\ket{v}\ket{a}\,=\,\ket{v'}\ket{a},
\end{equation}
where $v$ and $v'$ are adjacent and $a$ is the label of the edge $(v,v')$ as shown in Fig.~\ref{fig:SQWWH_graph6a}. $S^2=I$ as $S\ket{v'}\ket{a}\,=\,\ket{v}\ket{a}$ for all edges $(v,v')$. The $(+1)$-eigenvectors are
\begin{equation}
	\ket{\psi_{(v,v')}}\,=\,\frac{\ket{v}\ket{a}+\ket{v'}\ket{a}}{\sqrt 2},
\end{equation}
and there is a $(+1)$-eigenvector for each edge $(v,v')$. We are assuming that ${\psi_{(v,v')}}={\psi_{(v',v)}}$. Then
\begin{equation}\label{S}
	S\,=\,2\sum_{{(v,v')}\in E}\ket{\psi_{(v,v')}}\bra{\psi_{(v,v')}}-I.
\end{equation}
$S$ induces the red polygons of Fig.~\ref{fig:SQWWH_graph6b}.
After replacing each degree-$d$ vertex of $\Gamma$ by a $d$-clique, we obtain graph $\Gamma'$ of Fig.~\ref{fig:SQWWH_graph6b} on which the equivalent SQW is defined. The degree-5 vertex is converted into a 5-clique and the degree-3 vertex is converted into a 3-clique. The vertex labels of $\Gamma'$ have the form ``$v,j$'', where $v$ is the label of the vertex in the original graph and $j$ is the edge incident on $v$. With this notation, it is straightforward to check that the unitary and Hermitian operator that induces the red tessellation is $S$ given by Eq.~(\ref{S}), when we use vectors in uniform superposition.

Now, we can cast the evolution operator in the form demanded by the staggered model with Hamiltonians.
Since $iS=\textrm{e}^{i\pi S/2}$, the shift operator can be put in the form e$^{i\theta_1 H_1}$ with $\theta_1=\pi/2$ and $H_1=S$ modulo a global phase. If the coin is $\textrm{e}^{i\theta_0 H_0}$ and $H_0$ is an orthogonal reflection then any flip-flop coined QW on $\Gamma$ is equivalent to a SQW on $\Gamma'$ with evolution operator 
\begin{equation}
U=\textrm{e}^{i\frac{\pi}{2} S}\textrm{e}^{i\theta_0 H_0}.
\end{equation}
Operator $H_0$ induces the blue tessellation depicted in Fig.~\ref{fig:SQWWH_graph6b}. 

It is known that Grover's algorithm~\cite{Grover:1997a} can be described as a coined QW on the complete graph using a flip-flop shift operator and the Grover coin~\cite{Ambainis:2005, Portugal:Book}. Therefore, Grover's algorithm can also be reproduced by the SQW model~\cite{Por16}. Extensions of Grover's algorithm analyzed by Long~\textit{et al.}~\cite{LLZN99,LLZ02} and H\o{}yer~\cite{Hoy00} use operator
\begin{equation}\label{new_form}
	I-\left(1-\textrm{e}^{i\phi}\right) \ket{\psi}\bra{\psi},
\end{equation}
where $\ket{\psi}$ is the unit uniform superposition of the computational basis and $\phi$ is an angle, in place of the usual Grover operator $(I-2\ket{\psi}\bra{\psi})$. This kind of extension can be reproduced by SQW model with Hamiltonians because  $\textrm{e}^{i\theta_0 H_0}$ when $H_0$ is given by Eq.~(\ref{H_0}) can be written as
\begin{equation}
	I-\left(1-\textrm{e}^{2i\theta_0}\right)\sum_{k=0}^{m-1} \ket{\alpha_k}\bra{\alpha_k}
\end{equation}
modulo a global phase. We can choose values for $\theta_0$ and $m$ that reproduce Eq.~(\ref{new_form}).

\section{Conclusions}\label{sec5}

We have introduced an extension of the standard staggered QW model by using orthogonal reflections as Hamiltonians. Orthogonal reflections are local unitary operators in the sense that they respect the connections represented by the edges of a graph. Besides, orthogonal reflections are Hermitian by definition. This means that if $H_0$ is an orthogonal reflection of a graph $\Gamma$, then $U_0=\textrm{e}^{i\theta_0 H_0}$ is a local unitary operator associated with $\Gamma$. In order to define a nontrivial evolution operator, we need to employ a second orthogonal reflection $H_1$ of $\Gamma$. The generic form of the evolution operator of the SQW with Hamiltonians for 2-tessellable graphs is $U=\textrm{e}^{i\theta_1 H_1}\textrm{e}^{i\theta_0 H_0}$, where $\theta_0$ and $\theta_1$ are angles. This form is fitted for physical implementations in many physical systems, such as, cold atoms trapped in optical lattices~\cite{Greiner2002, Jaksch200552} and arrays of electromechanical resonators~\cite{1508.06984}.

We have obtained the wave function of SQWs with Hamiltonians on the line and analyzed the standard deviation of the probability distribution. For a localized initial condition at the origin, the maximum spread of the probability distribution for an evolution operator of the form $U=\textrm{e}^{i\theta H_1}\textrm{e}^{i\theta H_0}$ is obtained when $\theta=\pi/3$.

We have also characterized the class of coined QWs that are included in the SQW model with Hamiltonians and we have described how to convert those coined QWs on a graph $\Gamma$ into their equivalent formulation in terms of SQWs on an extended graph obtained from $\Gamma$ by replacing degree-$d$ vertices into $d$-cliques.

As a last remark, we call attention that recently it was shown numerically that searching one marked vertex using the original SQW on the two-dimensional square lattice has no speedup compared to classical search using random walks~\cite{FP16}. On the other hand, the SQW with Hamiltonians with $\theta=\pi/4$ is able to find the marked vertex after $O(\sqrt{N\log N})$ steps at least as fast as the equivalent algorithm using coined quantum walks~\cite{PD17}.

\section*{Acknowledgements}
RP acknowledges financial support from Faperj (grant n.~E-26/102.350/2013) and CNPq (grants n.~303406/2015-1, 4741\-43/2013-9) and also acknowledges useful discussions with Pascal Philipp and Stefan Boettcher. JKM acknowledges financial support from 
CNPq grant PDJ 165941/2014-6. MCO acknowledges support by FAPESP through the Research Center in Optics and Photonics (CePOF) and by CNPq.

%\bibliographystyle{unsrt}       % APS-like style for physics
%\bibliography{bib}   % name your BibTeX data base

\end{document}